\documentclass[12pt,prd,aps,amssymb,amsmath,tightenlines,showpacs,showkeys]{revtex4}

\usepackage{graphicx}

\newcommand{\cP}{{\cal P}}
\newcommand{\cT}{{\cal T}}
\newcommand{\cC}{{\cal C}}

\newcommand{\cPT}{{\cal PT}}

\begin{document}

\title{Extending $\cPT$ symmetry from Heisenberg algebra to E2 algebra}
\author{Carl~M.~Bender${}^1$}\email{cmb@wustl.edu}
\author{R.~J.~Kalveks${}^2$}\email{rudolph.kalveks09@imperial.ac.uk}

\affiliation{${}^1$Department of Physics, Washington University, St. Louis, MO
63130, USA \\
${}^2$Theoretical Physics, Imperial College London, London SW7 2AZ, UK}

\date{\today}

\begin{abstract}
The E2 algebra has three elements, $J$, $u$, and $v$, which satisfy the
commutation relations $[u,J]=iv$, $[v,J]=-iu$, $[u,v]=0$. We can construct the
Hamiltonian $H=J^2+gu$, where $g$ is a real parameter, from these elements. This
Hamiltonian is Hermitian and consequently it has real eigenvalues. However, we
can also construct the $\cPT$ symmetric and non-Hermitian Hamiltonian $H=J^2+ig
u$, where again $g$ is real. As in the case of $\cPT$-symmetric Hamiltonians
constructed from the elements $x$ and $p$ of the Heisenberg algebra, there are
two regions in parameter space for this $\cPT$-symmetric Hamiltonian, a region
of unbroken $\cPT$ symmetry in which all the eigenvalues are real and a region
of broken $\cPT$ symmetry in which some of the eigenvalues are complex. The two
regions are separated by a critical value of $g$.
\end{abstract}


\maketitle

\section{Introduction}
\label{s1}

The Heisenberg algebra for a quantum mechanical system having one degree of
freedom consists of three elements: the coordinate operator $x$, the momentum
operator $p$, and the unit element ${\bf 1}$. These three elements obey the
single commutation relation
\begin{equation}
[x,p]={\bf 1}i
\label{e1}
\end{equation}
This algebra possesses two independent discrete symmetries under which the
commutation relation (\ref{e1}) remains invariant. The first symmetry, called
parity (space reflection), is represented by the linear operator $\cP$, where
$\cP^2={\bf 1}$. Under the action of $\cP$ both $x$ and $p$ change sign:
\begin{equation}
\cP x\cP=-x,\qquad\cP p\cP=-p.
\label{e2}
\end{equation}
The second symmetry, called time reversal, is represented by the antilinear
operator $\cT$, where $\cT^2={\bf 1}$. Under the action of $\cT$ both $p$ and
$i$ change sign, but $x$ does not:
\begin{equation}
\cT p\cT=-p,\qquad\cT i\cT=-i,\qquad\cT x\cT=x.
\label{e3}
\end{equation}

In quantum mechanics the Hamiltonian operator $H$ is expressed in terms of the
operators $x$ and $p$: $H=H(x,p)$. It is conventional to require that the
Hamiltonian be Hermitian so that the eigenvalues of $H$ are real. However, in
1998 it was shown that the Hamiltonian need not be Hermitian for the eigenvalues
to be real \cite{R1,R2}. In that paper the family of non-Hermitian Hamiltonians
\begin{equation}
H=p^2+x^2(ix)^\epsilon,
\label{e4}
\end{equation}
was introduced, and it was shown that the eigenvalues of these $\cPT$-symmetric
Hamiltonians are real when the parameter $\epsilon\geq0$. A rigorous proof of
spectral reality is given in Refs.~\cite{R3,R4}. The parametric range $\epsilon
\geq0$ is referred to as a region of unbroken $\cPT$ symmetry; in this region
all the eigenfunctions of $H$ are also eigenfunctions of the $\cPT$ operator. In
the parametric region $\epsilon<0$ some of the eigenvalues are complex; this
range of $\epsilon$ is said to be a region of broken $\cPT$ symmetry.

A second conventional reason for requiring that the Hamiltonian $H$ be Hermitian
is that $H$ determines the time evolution of the theory, and if $H$ is
Hermitian, then the time evolution is unitary (probability conserving). However,
in 2002 it was shown that if the $\cPT$ symmetry of a non-Hermitian Hamiltonian
is unbroken, then the time evolution is unitary \cite{R5}.

In general, Hermitian Hamiltonians differ from $\cPT$-symmetric Hamiltonians in
that the spectrum of a Hermitian Hamiltonian is always real while the spectrum
of a non-Hermitian $\cPT$-symmetric Hamiltonian often has a parametric region of
unbroken $\cPT$ symmetry where the eigenvalues are all real and a region of
broken $\cPT$ symmetry where some of the eigenvalues are complex. The boundary 
between these two regions is a phase transition, and this phase transition has
recently been observed in several different laboratory experiments
\cite{R6,R7,R8,R9}.

In this paper we consider the algebra E2, which is more complicated than the
Heisenberg algebra. This algebra has three elements, which are designated $J$,
$u$, and $v$, and these elements obey the commutation relations
\begin{equation}
[u,J]=iv,\qquad [v,J]=-iu,\qquad [u,v]=0. 
\label{e5}
\end{equation}
This operator algebra arises when one considers a two-dimensional quantum system
restricted to a ring of radius $r$. We can represent the operators $J$, $u$, and
$v$ in polar form as
\begin{equation}
J=-i\frac{\partial}{\partial\theta},\qquad u=\sin\theta,\qquad v=\cos\theta,
\label{e6}
\end{equation}
where we have taken $r=1$. The E2 algebra is more complicated than the
Heisenberg algebra (\ref{e1}), but it reduces to the Heisenberg algebra in the
limit as $r\to\infty$ \cite{R10}.

Like the Heisenberg algebra, the E2 algebra is separately invariant under each
of two different symmetries, parity $\cP$ and time reversal $\cT$. We define a
parity transformation as a reflection through the center of the ring. Thus, a
point on the ring is mapped to a point on the opposite side of the ring such
that $\theta\to\theta+\pi$. Then, under a parity transformation
\begin{equation}
\cP J\cP=J,\qquad\cP u\cP=-u,\qquad \cP v\cP=-v,
\label{e7}
\end{equation}
which clearly leaves (\ref{e5}) invariant \cite{R11}. Time reversal changes the
sign of $i$, and thus its effect is to reverse the sign of $J$ but to leave $u$
and $v$ invariant. Thus, the time-reversal transformation
\begin{equation}
\cT J\cT=-J,\qquad\cT u\cT=u,\qquad \cT v\cT=v
\label{e8}
\end{equation}
also leaves the E2 algebra (\ref{e5}) invariant.

This paper is organized very simply: In Sec.~\ref{s2} we construct a Hamiltonian
in terms of the elements of the E2 algebra. This Hamiltonian contains a
coupling-constant parameter $g$; if $g$ is real, the Hamiltonian is Hermitian,
and if $g$ is imaginary, the Hamiltonian is non-Hermitian but $\cPT$ symmetric.
We show that for real $g$ the eigenvalues are all real and that if $g$ is
imaginary, there are regions of broken and unbroken $\cPT$ symmetry. Finally, in
Sec.~\ref{s3} we give some brief concluding remarks and discuss possible future
directions for research.

\section{Hermitian and $\cPT$-Symmetric Hamiltonians constructed from the
elements of E2}
\label{s2}

The operators $J$, $u$, and $v$ are Hermitian, and thus it is easy to construct
a Hermitian Hamiltonian in terms of these operators. One such Hamiltonian is 
\begin{equation}
H=J^2+gv,
\label{e9}
\end{equation}
where $g$ is a real parameter. For this Hamiltonian, the Schr\"odinger
eigenvalue differential equation takes the form
\begin{equation}
-\psi''(\theta)+g\cos(\theta)\psi(\theta)=E\psi(\theta),
\label{e10}
\end{equation}
which is the Mathieu equation \cite{R12}. This is the equation for a quantum
pendulum.

To find the eigenvalues $E$ of (\ref{e10}) we must impose boundary conditions.
The simplest such boundary conditions express the bosonic requirement that the
eigenfunctions be single valued on the ring:
\begin{equation}
\psi(\theta+2\pi)=\psi(\theta).
\label{e11}
\end{equation}
However, instead of imposing the $2\pi$-periodic boundary conditions in
(\ref{e11}), we can also impose the fermionic boundary requirement that the
eigenfunctions be $2\pi$ {\it anti}-periodic:
\begin{equation}
\psi(\theta+2\pi)=-\psi(\theta).
\label{e12}
\end{equation}
Because these boundary conditions are homogeneous and the Schr\"odinger equation
(\ref{e10}) is symmetric under $\theta\to-\theta$, the eigenfunctions will be
either odd or even in $\theta$.

Let us examine first the simple case $g=0$. The general solution to (\ref{e10})
for this case is
\begin{equation}
\psi(x)=A\sin\left(\sqrt{E}\theta\right)+B\cos\left(\sqrt{E}\theta\right).
\label{e13}
\end{equation}
Thus, there are four sets of eigenvalues: For odd bosonic eigenfunctions
$B=0$, and we get
\begin{equation}
E_n=n^2\quad(n=1,~2,~3,~\ldots);
\label{e14}
\end{equation}
for even bosonic eigenfunctions $A=0$, and we get
\begin{equation}
E_n=n^2\quad(n=0,~1,~2,~3,~\ldots);
\label{e15}
\end{equation}
for odd fermionic eigenfunctions $B=0$, and we get
\begin{equation}
E_n=\textstyle{\frac{1}{4}}n^2\quad(n=1,~3,~5,~7,~\ldots);
\label{e16}
\end{equation}
for even fermionic eigenfunctions $A=0$, and we get
\begin{equation}
E_n=\textstyle{\frac{1}{4}}n^2\quad(n=1,~3,~5,~7,~\ldots).
\label{e17}
\end{equation}

\begin{figure}
\begin{center}
\includegraphics[scale=0.35, bb=0 0 1000 607]{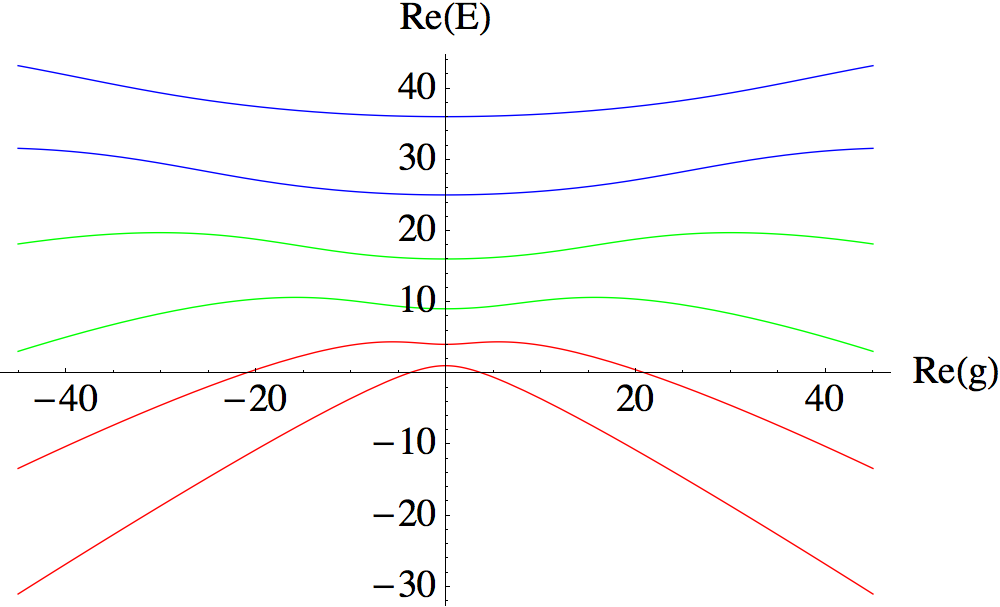}
\end{center}
\caption{Odd bosonic eigenvalues for the Schr\"odinger equation (\ref{e7})
plotted as a function of real $g$. The spectrum for $g=0$ is given in
(\ref{e14}).}
\label{F1}
\end{figure}

\begin{figure}
\begin{center}
\includegraphics[scale=0.35, bb=0 0 1000 600]{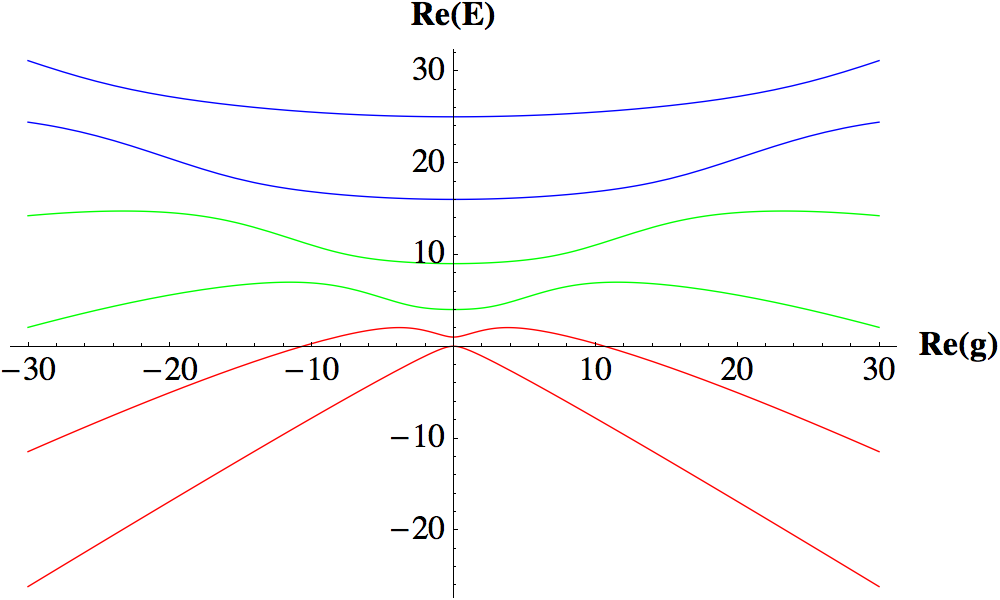}
\end{center}
\caption{Even bosonic eigenvalues for the Schr\"odinger equation (\ref{e7}) 
plotted as a function of real $g$. The spectrum for $g=0$ is given in 
(\ref{e15}).}
\label{F2}
\end{figure}

For $g\neq0$ we use Mathematica to plot the eigenvalues as functions of $g$. The
odd bosonic eigenvalues are shown in Fig.~\ref{F1} and the even bosonic
eigenvalues are shown in Fig.~\ref{F2}. Note that because the Hamiltonian is
Hermitian, the eigenvalues are all real.

\begin{figure}
\begin{center}
\includegraphics[scale=0.22, bb=0 0 2027 644]{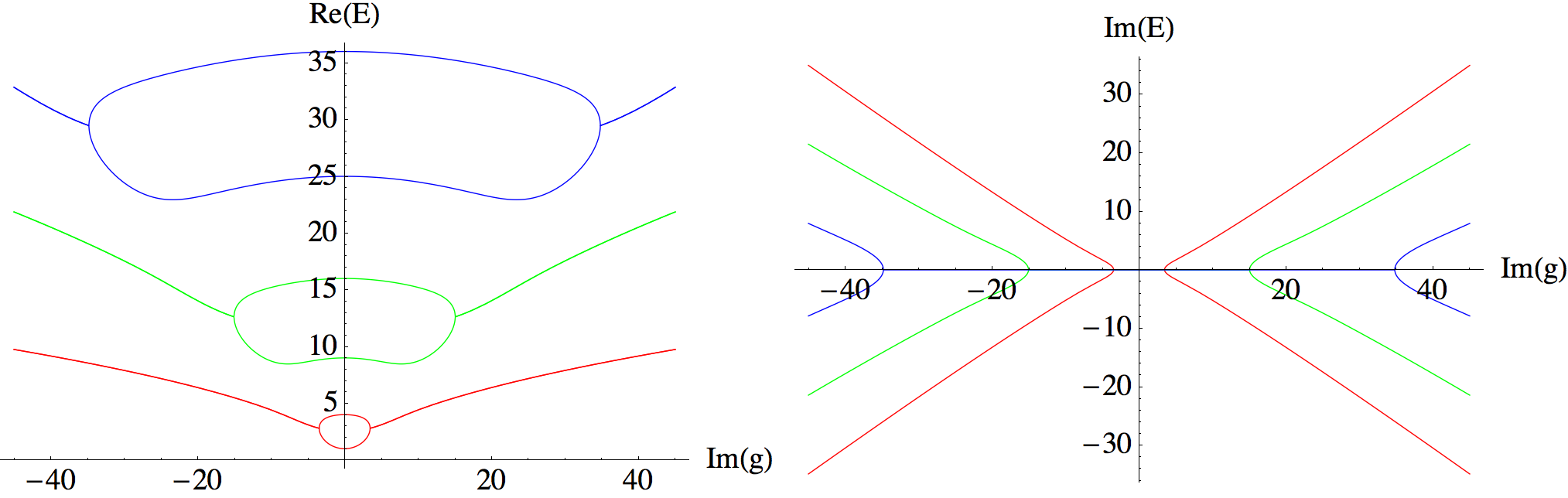}
\end{center}
\caption{Odd bosonic eigenvalues for the $\cPT$-symmetric Hamiltonian (\ref{e9})
in which the parameter $g$ is pure imaginary. The eigenvalues are plotted as
functions of ${\rm Im}\,g$. The real (imaginary) parts of the eigenvalues are
shown in the left (right) panel. Observe that the eigenvalues are all real when
$-3.4645<{\rm Im}\,g<3.4645$; this is the region of unbroken $\cPT$ symmetry.
There is an infinite sequence of critical points; the next critical points are
at ${\rm Im}\,g=\pm15.0485$ and at $\pm34.7994$.}
\label{F3}
\end{figure}

\begin{figure}
\begin{center}
\includegraphics[scale=0.22, bb=0 0 2027 630]{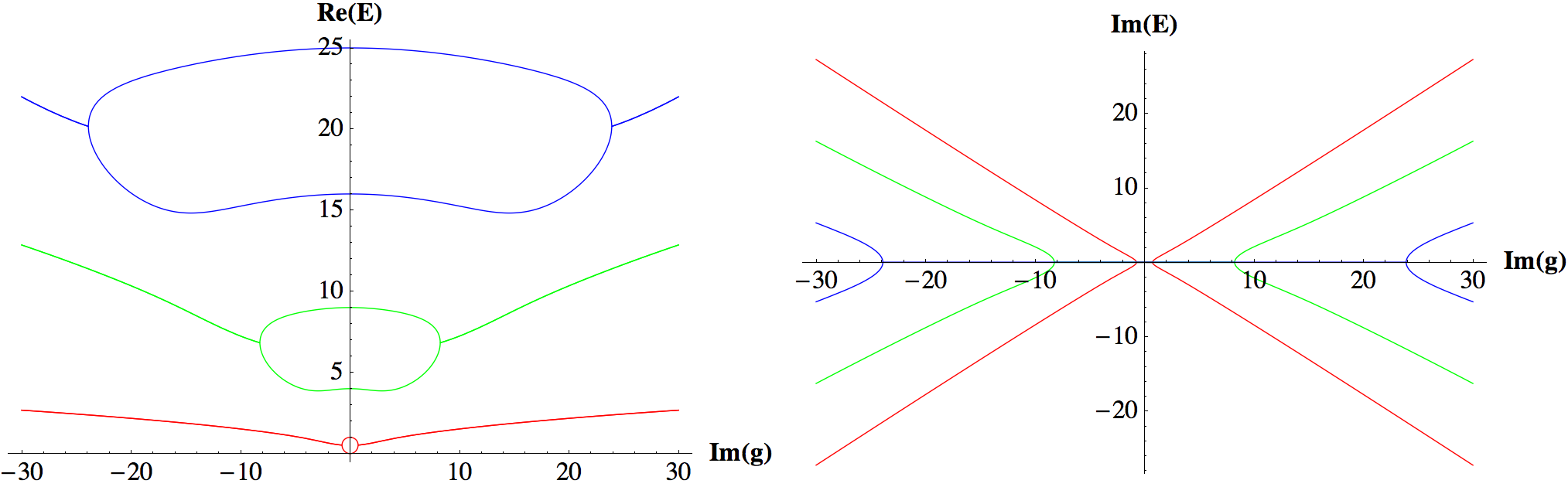}
\end{center}
\caption{Even bosonic eigenvalues for the $\cPT$-symmetric Hamiltonian
(\ref{e9}) plotted as functions of ${\rm Im}\,g$. The real (imaginary) parts of
the eigenvalues are shown in the left (right) panel. The eigenvalues are all
real when $-0.7344<{\rm Im}\,g<0.7344$; this is the region of unbroken $\cPT$
symmetry. In the regions of broken $\cPT$ symmetry there is an infinite sequence
of critical points; the next critical points are at ${\rm Im}\,g=\pm8.2356$ and
at $\pm23.9030$.}
\label{F4}
\end{figure}

Now let us see what happens if we take the parameter $g$ in the Hamiltonian
(\ref{e9}) to be pure imaginary. This choice of $g$ makes the Hamiltonian
non-Hermitian but $\cPT$-symmetric. For ${\rm Im}\,g\neq0$ we again use
Mathematica to plot the eigenvalues as functions of ${\rm Im}\,g$. Because the
Hamiltonian is no longer Hermitian, some of the eigenvalues are complex. The
real (left panel) and imaginary (right panel) parts of the eigenvalues are shown
in Figs.~\ref{F3} and \ref{F4}, with the odd bosonic eigenvalues in
Fig.~\ref{F3} and the even bosonic eigenvalues in Fig.~\ref{F4}. The key feature
of the spectrum is that all of the eigenvalues are {\it real} if ${\rm Im}\,g$
lies between the critical values $-3.4645$ and $3.4645$ for the odd bosonic
eigenvalues and between $-0.7344$ and $0.7344$ for the even bosonic eigenvalues.
This is the region of unbroken $\cPT$ symmetry. As $|{\rm Im}\,g|$ increases
past these critical points, the lowest two eigenvalues become degenerate and
move into the complex plane as a complex-conjugate pair. Thus, we have entered
the regions of broken $\cPT$ symmetry. In fact, there is an infinite sequence of
critical points: The next two lowest pairs of eigenvalues become degenerate and
move into the complex plane at the critical points $\pm15.0485$ and $\pm34.7994$
for the odd bosonic eigenvalues and at $\pm8.2356$ and $\pm23.9030$ for the even
bosonic eigenvalues.

\begin{figure}
\begin{center}
\includegraphics[scale=0.22, bb=0 0 2027 644]{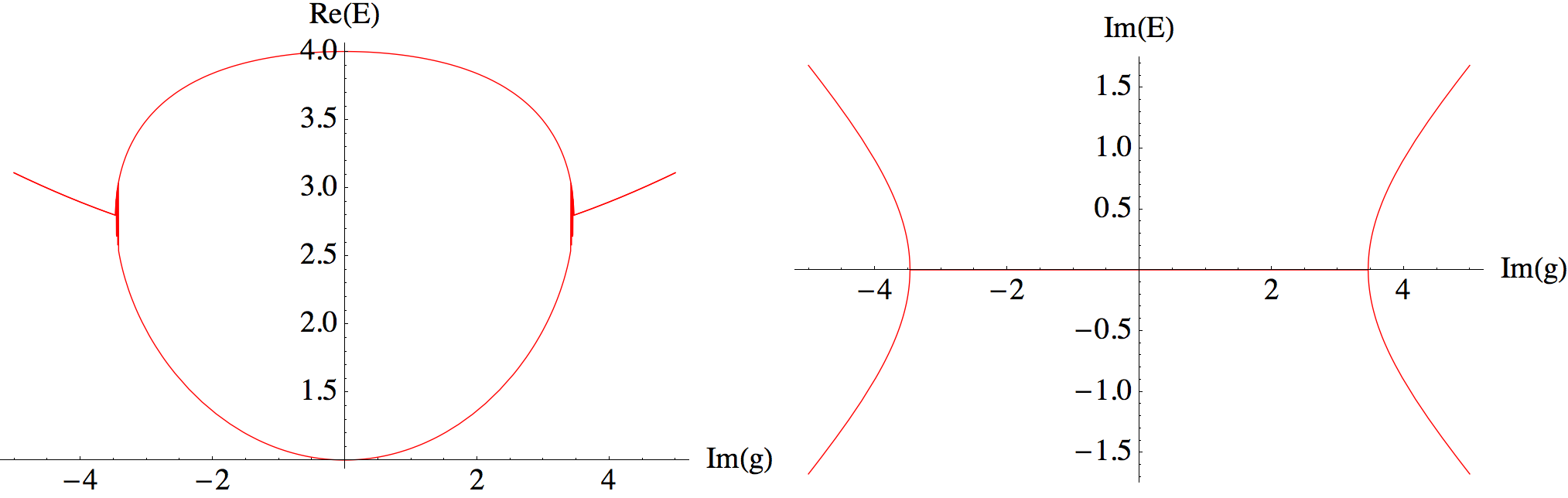}
\end{center}
\caption{Blow-up of the region near the critical points at ${\rm Im}\,g=\pm
3.4645$ on Fig.~\ref{F3}. The imaginary parts of the two lowest energy levels
vanish until ${\rm Im}\,g$ passes a critical point. At this point the two energy
levels become degenerate and ${\rm Im}\,E(g)$ for each energy level suddenly
makes a $90^\circ$ turn. This is the typical behavior of a function near a
square-root singularity.}
\label{F5}
\end{figure}

\begin{figure}
\begin{center}
\includegraphics[scale=0.22, bb=0 0 2027 628]{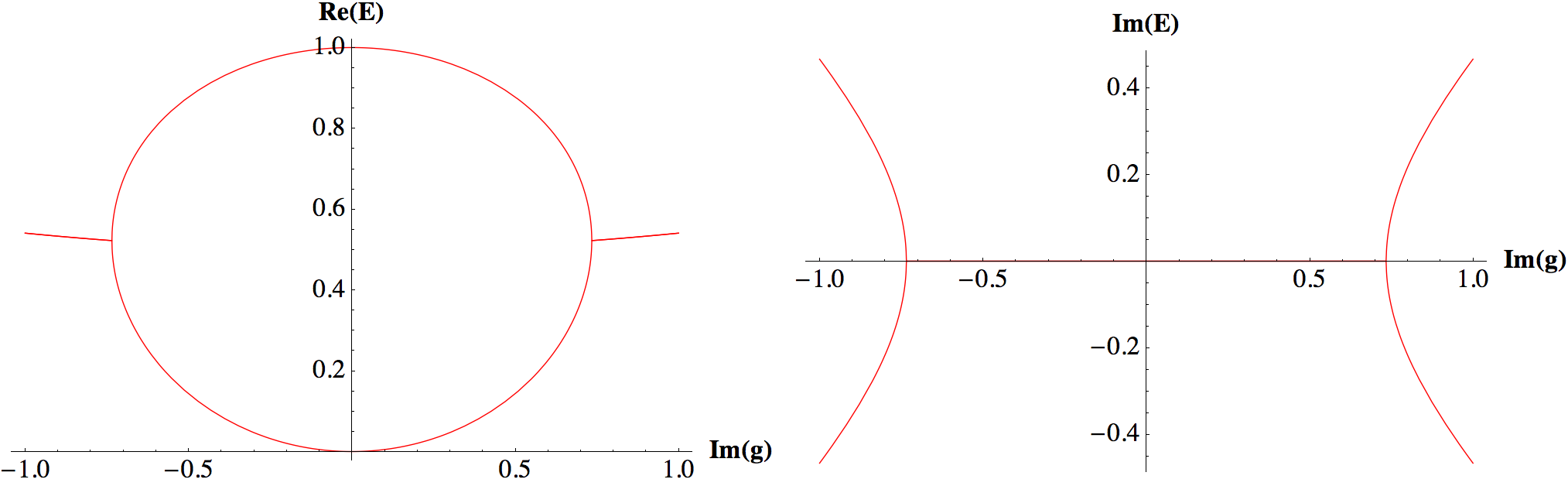}
\end{center}
\caption{Blow-up of the region near the critical points at ${\rm Im}\,g=0.7344$
on Fig.~\ref{F4}. As in Fig.~\ref{F5}, the imaginary part of the energies of the
two lowest states is 0 until ${\rm Im}\,g$ reaches a critical point. At this
point the energy levels merge and become a complex-conjugate pair.}
\label{F6}
\end{figure}

Figures~\ref{F5} and \ref{F6} give detailed plot of the transition from
unbroken to broken $\cPT$ symmetry. Observe that the eigenvalues become
degenerate in pairs and that the real and imaginary parts of the eigenvalues
make $90^\circ$ turns at the critical points. This is a clear indication that
the critical points are square-root branch points.

\section{Concluding remarks}
\label{s3}

We have shown in this paper that the well studied properties of $\cPT$-symmetric
quantum mechanical Hamiltonians that are constructed from the elements of the 
Heisenberg algebra extend to Hamiltonians that are constructed from the elements
of the E2 algebra. Both algebras are individually invariant under parity
reflection $\cP$ and under time reversal $\cT$. If a Hamiltonian that is
constructed from the elements of either of these algebras is Hermitian, then its
eigenvalues are all real. However, if the Hamiltonian is non-Hermitian and
$\cPT$ symmetric, then there may be regions of unbroken and unbroken $\cPT$
symmetry.

\begin{figure}
\begin{center}
\includegraphics[scale=0.22, bb=0 0 2027 636]{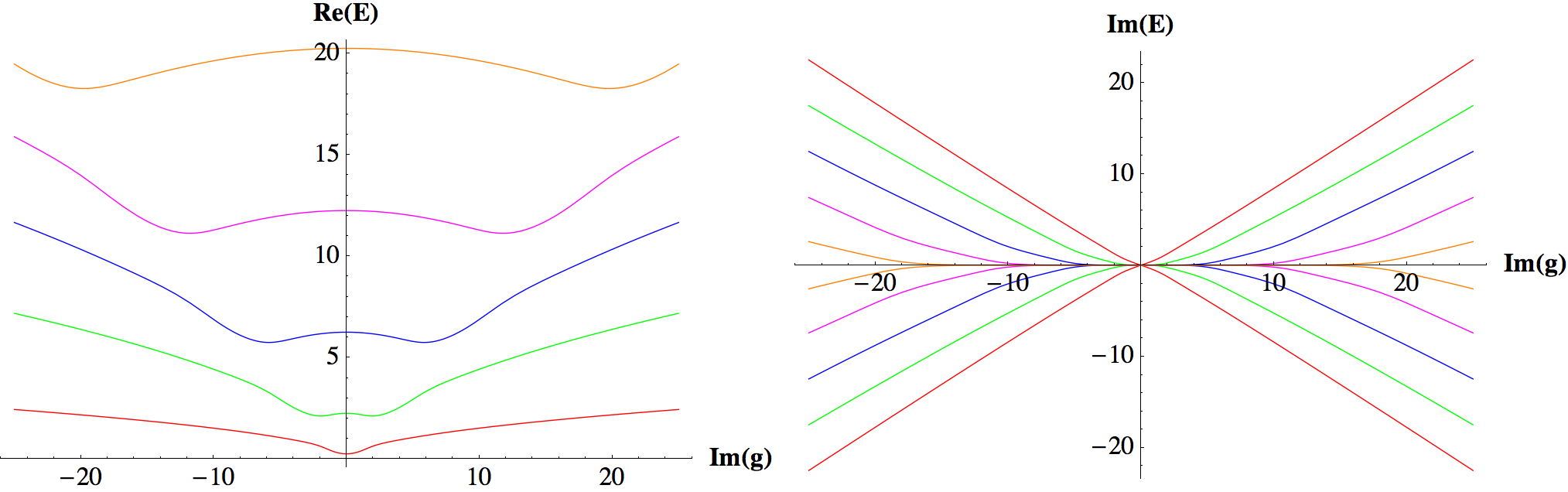}
\end{center}
\caption{Fermionic eigenvalues for the $\cPT$-symmetric Hamiltonian (\ref{e9})
plotted as functions of ${\rm Im}\,g$. The real (imaginary) parts of the
eigenvalues are shown in the left (right) panel. The eigenvalues are all complex
when ${\rm Im}\,g\neq0$; thus, there is no region of unbroken $\cPT$ symmetry.
The eigenvalues for both even and odd eigenfunctions are shown; the even and
odd eigenvalues form complex-conjugate pairs. Five pairs of eigenvalues are
shown in the figure.}
\label{F7}
\end{figure}

It is interesting that for the fermionic eigenvalues of the Hamiltonian
(\ref{e9}) there is no region of unbroken $\cPT$ symmetry; that is, the
eigenvalues are all complex when $g$ is nonzero and purely imaginary, as shown
in Fig.~\ref{F7}. Note from (\ref{e16}) and (\ref{e17}) that the odd and even
fermionic eigenvalues are degenerate when $g=0$. When ${\rm Im}\,g\neq0$, this 
degeneracy is split, and the eigenvalues become complex-conjugate pairs. Thus,
the condition of $\cPT$ symmetry seems to exclude real fermionic eigenvalues.
The nonexistence of fermionic eigenvalues was already observed in earlier
studies of $\cPT$-symmetric crystal lattices \cite{R13,R14}. In these studies
the discriminant was calculated as a function of the energy $E$. For Hermitian
periodic potentials the discriminant $D(E)$ is a smooth, real oscillatory
function of $E$. The inequality $|D(E)|<2$ identifies a band of allowed energies
and the inequality $|D(E)|>2$ defines a gap of forbidden energies. At the band
edge $D(E)=2$ the eigenfunction is bosonic ($2\pi$ periodic) and at the band
edge $D(E)=-2$ the eigenfunction is fermionic ($2\pi$ antiperiodic). In
Refs.~\cite{R13,R14} it was found that for $\cPT$-symmetric periodic potentials
half of the gaps disappeared and at the band edges the eigenfunction is only
bosonic.

The current work complements the work in Refs.~\cite{R13,R14}. Rather than
calculating the discriminant as a function of the energy $E$, we have calculated
the energies of periodic bosonic and antiperiodic fermionic eigenfunctions as
functions of the coupling constant $g$. We find while bosonic eigenfunctions
have a region of unbroken $\cPT$ symmetry, fermionic eigenfunctions do not
have such a region.

There are some natural continuations of this research. It is important to verify
that the higher algebras E3, E4, and so on, are also invariant under $\cP$ and
$\cT$ transformations and that non-Hermitian $\cPT$-symmetric Hamiltonians
constructed from the elements of these algebras have regions of unbroken and
broken $\cPT$ symmetry. Furthermore, it would be most interesting to calculate 
the $\cC$ operator \cite{R5,R15,R16} in the unbroken $\cPT$-symmetric regions.

\begin{acknowledgments}
We thank C.~J.~Isham for many inspiring discussions. CMB is grateful to the
U.S.~Department of Energy for financial support. Mathematica 7 was used to
generate the figures in this paper.
\end{acknowledgments}


\begin{thebibliography}{99}
\bibitem{R11} There are other acceptable definitions of parity reflection. For
example, if we define such a reflection to have the effect $\theta\to-\theta$,
then the effect of $\cP$ is to make the replacements $J\to-J$, $u\to-u$, and
$v\to v$. The E2 algebra is invariant under this transformation. However, if we
use this definition instead of the definition (\ref{e7}) used in this paper, the
results reported here remain unchanged.
\end{thebibliography}
\end{document}